\documentclass[a4paper, 12pt]{article}
\def\be{\begin{equation}}
\def\ee{\end{equation}}
\def\s{\scriptscriptstyle}
\def\ba{\begin{eqnarray}}
\def\ea{\end{eqnarray}}
\def\1{\hskip1pt}
\def\2{\hskip2pt}
\def\5{\hskip5pt}
\def\v{\mbox{vol\1}}

\renewcommand{\l}[1]{\mbox{$\scriptstyle \lambda_{\s #1}$}}
\newcommand{\kd}[1]{\mbox{ker\1$d_{\s #1}$}}

\def\i{\mbox{Im\1}}
\newcommand{\dd}[1]{\mbox{det}^{\prime}\1(d^{\dagger}_{\s #1}
d^{\phantom{\dagger}}_{\s #1})}
\newcommand{\df}[1]{\mbox{det\1}(\phi^{\dagger}_{\s #1}
		\phi^{\phantom{\dagger}}_{\s #1})}
\renewcommand{\o}[1]{\mbox{$\Omega^{\scriptstyle #1}$(M)}}
\newcommand{\h}[1]{\mbox{$\mathcal{H}^{\scriptstyle #1}$(M)}}
\renewcommand{\H}[1]{\mbox{H$_{\mbox{\tiny dR}}^{\scriptstyle #1}$(M)}}
\renewcommand{\r}{\2 R $\longleftrightarrow 1/\mbox{R\5}$}
\def\+{^{\s 1/2 \2}}
\def\-{^{\s -1/2 \2}}
\begin{document}
\begin{center}
{\huge\textbf{Abelian Duality \\}}
\vspace {6mm}
\renewcommand{\thefootnote}{$\ast$}
{\large Emil M. Prodanov\footnote {Supported by FORBAIRT scientific research
program and Trinity College, Dublin.} and Siddhartha Sen} \\
\vspace {2mm}
{\it School of Mathematics, Trinity College, Dublin 2, Ireland, \\
E-Mail: \hskip 2pt prodanov@maths.tcd.ie, \hskip 2pt sen@maths.tcd.ie}
\end{center}
\vspace{4mm}
\begin{abstract}
We show that on three--dimensional Riemannian manifolds without boundaries and 
with trivial first real de Rham cohomology group (and in no other dimensions) 
scalar field theory and Maxwell theory are equivalent: the ratio of the 
partition functions is given by the Ray--Singer torsion of the manifold. 
At the level of interaction with external currents, the equivalence persists 
provided there is a fixed relation between the charges and the currents. 
\end{abstract}
\normalsize
\newpage

\noindent
There has been recent interest in relating different theories and establishing
their equivalence. Common to all applications of the different aspects of the 
notion of duality is the observation that when two different theories are dual
to each other, then either the manifolds are changed or the fields and the 
coupling constants are related. \\
In this paper we re-examine two simple systems --- scalar field theory and 
Maxwell theory on three--dimensional Riemannian manifolds without boundaries 
and with trivial $\pi_{\s 1}$(M). We show the equivalence 
between these theories and we give the condition which must be satisfied by 
the charges and the external currents in order that the equivalence persists 
at the level of interactions. This is done by a direct calculation of the 
partition function of each theory paying particular attention to the fine 
structure of the zero--mode sector. In the spirit of Schwarz's method of 
invariant integration~\cite{s2} we show that the ratio of the partition 
functions of the theories is equal to the square of the partition function of 
Chern--Simons theory (or the partition function of BF theory, that is, 
U(1)$\times$U(1) Chern--Simons theory with purely off-diagonal coupling). Such 
equivalence between a scalar and vector theory is a novel form of duality 
which we call Abelian Duality. We show that when the coupling constants 
(overall scaling factors) are related as \r, then this Abelian Duality
transforms into \r duality. In this case the ratio of the partition functions 
is given by a topological invariant --- the Ray--Singer torsion of the 
manifold. We show how our results can be obtained by Schwarz's resolvent 
method~\cite{s1} and we use a resolvent generated by the de Rham complex to 
comment on possibilities of equivalence between the theories in other 
dimensions. In our considerations we use zeta--regularised determinants.

\vskip1cm
\noindent
We will consider Riemannian manifolds M. All operators entering our 
theory can be described by the following diagram:
\vskip1cm
\begin{picture}(300, 100)\thinlines
\put(105, 75){\parbox{45pt}{\o{p}}}
\put(152, 79){\vector(1,0){50}}
\put(212, 75){\parbox{50pt}{\o{p+1}}}
\put(118, 40){\vector(0,1){30}} 
\put(230, 70){\vector(0,-1){30}}
\put(100, 28){\parbox{50pt}{\o{m-p}}}
\put(152, 32){\vector(1,0){50}}
\put(208, 28){\parbox{63pt}{\o{m-p-1}}}
\put(160, 38){\parbox{50pt}{$\scriptstyle d_{\s m-p-1}^{\dagger}$}}
\put(170, 82){\parbox{50pt}{$\scriptstyle d_{\s p}$}}
\put(122, 52){\parbox{10pt}{$\scriptstyle \ast$}}
\put(234, 52){\parbox{10pt}{$\scriptstyle \ast$}}
\put(358, 28){\parbox{20pt}{(1)}}
\end{picture}
\addtocounter{equation}{1}

\noindent
where $m=$ dim M. The case of interest will be a three-dimensional manifold. \\
We will first calculate the partition function of free Maxwell theory:
\ba
Z_{\s 1}(\lambda_{\s 1}) \5 & = & \5 
\int\limits_{\Omega^{\s 1} (M)}\!\!\!\!\mathcal{D}A \5 \1
e^{\2-i \1 \l{1} \int \!\! d^{\s 3}x \2 \sqrt{g} \2 \2
F_{\s \mu \nu}\1 F^{\s \mu \nu}} \5 = \5 
\int\limits_{\Omega^{\s 1} (M)}\!\!\!\!\mathcal{D}A \5 \1 
e^{\2-i \1 \l{1} \int d_{\s 1}A \2 \wedge\2 \ast\2 
d_{\s 1}A} \nonumber \\ 
& = & \5 \int\limits_{\Omega^{\s 1} (M)}\!\!\!\!\mathcal{D}A \5 \1 
e^{\2-i \1 \l{1} \2 \langle d_{\s 1}A, \2\2 d_{\s 1}A \rangle}
\5 = \5 \int\limits_{\Omega^{\s 1} (M)}\!\!\!\!\mathcal{D}A \5 \1
e^{\2-i \1 \l{1} \2 \langle A, \2\2 d^{\dagger}_{\s 1} 
d^{\phantom{\dagger}}_{\s 1}A \rangle}.
\ea
Note that here we have written \2 $F$ \2 as \2 $d_{\s 1}A$.\5  
The Maxwell equation \2 $d_{\s 2}F=0$ \2 implies that \2 $F$ \2 is an element 
in \2 \H{2}.\5 $A$ \2 is a one-form and therefore \2 $d_{\s 1}A=0$ \2 in 
\2 \H{2}, \2 that is, if \2 $F=d_{\s 1}A$ \2 then the equivalence class \2
$[F]$ \2 is zero in \2 \H{2} \2 (i.e. \2 $[F] = [F^{\prime}] 
\Longleftrightarrow F = F^{\prime} + d_{\s 1}A$). When the second de Rham 
cohomology group is trivial, then \2 $F=d_{\s 1}A$ \2 is valid globally. In 
three dimensions \2 \H{2} \2 is isomorphic to \2 \H{1} \2 due to Hodge 
duality and \2 \H{1} \2 being trivial means that the first homotopy 
group \2 $\pi_{\s 1}$(M) \2 is trivial (then \2 $F=d_{\s 1}A$ \2 globally). If
\2 $\pi_{\s 1}$(M)\2 is non-trivial, then \2 $F=d_{\s 1}A$ \2 is valid only on 
contractible regions of the manifold.  \\
We will show the equivalence between Maxwell theory and scalar field theory 
for the case when $F$ could be written as $d_{\s 1}A$ globally, that is, for 
homology 3-spheres $\Bigl($i.e. manifolds with trivial first real de Rham 
cohomology group, e.g. {\bf S}$^{\s 3}$ or the lens spaces L$(p, q)\Bigr)$. 
For the general case the duality is more subtle and we would like to refer 
the reader to~\cite{w} where Witten has shown how to pass from scalar field
theory to Maxwell theory and vice versa in two and three dimensions. \\
We need manifolds without boundaries because we want to integrate by parts to 
bring the differential operator $d_{\s 1}$ on the other side of the scalar 
product and at the same time not to be bothered about boundary terms. \\ 
We need Riemannian manifolds, because we are dealing with partition 
functions. \\
The integral is over the space of all 1-forms \o{1}. We can decompose the 
space of all 1-forms as a direct sum of the kernel of the operator entering 
the partition function and its orthogonal complement:
\be
\o{1} = 
\kd{1} \oplus (\kd{1})^{\perp}.
\ee
Therefore
\be
Z_{\s 1}(\lambda_{\s 1}) =  \v(\kd{1})
\!\!\!\!\int\limits_{(\mbox{\scriptsize ker\2} d_{\s 1})^{\perp}}
\!\!\!\!\mathcal{D}A\5\1
e^{\2-i \2 \l{1} \2 \langle A, \2\2 d^{\dagger}_{\s 1} 
d^{\phantom{\dagger}}_{\s 1}A \rangle}  = 
\v(\kd{1}) \2 \2 
\mbox{det}^{\prime}\1\Bigl(\frac{i \l{1}}{\pi}\2 d^{\dagger}_{\s 1}
d^{\phantom{\dagger}}_{\s 1}\Bigr)\- .
\ee
The partition function is thus an ill-defined quantity --- the determinant and 
the volume factor are infinite. Our calculations will be formal. With a 
zeta--regularization technique we can make the determinant finite. We can also
asssume that an appropriate normalization is chosen in such way that the 
divergency of the volume factor is absorbed. This will make the partition 
function finite. \\
Note that \v(\kd{1}) is nothing else but the Faddeev--Popov ghost determinant 
times the ghost--for--ghost determinant. To see that, let us calculate the 
same partition function using the method of invariant integration~\cite{s2}. 
In other words we will exploit the gauge symmetry of the theory to restrict 
the integration over \o{1} to integration over a lower-dimensional space --- 
the space of the orbits of the group of gauge transformation. \\
The stabilizer of the group of gauge transformations $A \longrightarrow
A + d_{\s 0}\o{0}$ consists of those elements of $\o{0}$ 
for which $d_{\s 0}\o{0} = 0$, that is, the constant functions.
In order to pick one representative of each equivalence class $[A]$, we impose
a gauge condition, that is, we intersect the space of the orbits of the group
of gauge transformations in the space of all 1-forms by a hyperplane defined 
by those $A$'s, for which $\partial_{\mu} A^{\mu} = 0$, i.e. 
$d^{\dagger}_{\s 0}A  = 0$. The integration is then performed over this 
hyperplane. We can always make sure that the element at each intersection 
point of a group orbit with this hyperplane is not a zero--mode of the 
operator. Thus, the Faddeev--Popov trick not only restricts the gauge freedom,
but also isolates the zero--modes of the operator. The Faddeev--Popov 
determinant is a delta function of the gauge-fixing condition (Lorentz gauge 
in our case). In addition we must multiply by the volume of each 
orbit in order to preserve the value of the partition function. That is, we 
must divide by the volume of the stabilizer of the group at each point (we 
have assumed that the volume of the group of gauge transformations is 
normalized to one). This volume factor is the ghost--for--ghost 
determinant --- as the Faddeev--Popov determinant is not 
finite~\cite{da} --- $d_{\s 0}$ itself has zero--modes (it vanishes on the 
constant functions). So we need an analogue of the gauge-fixing 
condition --- this time for the ghosts, not for the fields. \\
Assuming that all stabilizers are conjugate, we get:
\be
Z_{\s 1}(\lambda_{\s 1}) \5  =  \5 \frac
{1}{\v(\kd{0})}
\int\limits_{\raise 2pt\hbox{\scriptsize $\Omega^{\s 1}(M)$}/ 
\lower 2pt\hbox{$\scriptstyle d_{\s 0}$}}
\!\!\!\!\mathcal{D}[A]\5\1
e^{\2-i \2 \l{1} \2 \langle A, \2\2 d^{\dagger}_{\s 1} 
d^{\phantom{\dagger}}_{\s 1}A \rangle}\5
\mbox{det$^\prime$\2}(d_{\s 0}^{\dagger}d_{\s 0}^{\phantom{\dagger}})^{\s 1/2}.
\ee
The stabilizer of the group of gauge transformations consists of the constant
functions, that is, the stabilizer is the real line. The real line can be
canonically identified with the zeroth de Rham cohomology group \H{0}. The 
projection map \kd{q} $\longrightarrow$ \H{q} induces the 
isomorphism~\cite{da}:
\be
\phi_{\s q}: \mbox{\h{q}}\longrightarrow \mbox{\H{q}}
\ee
where \h{q} is the space of harmonic q-forms. Therefore:
\be
\v\Bigl(\mbox{\h{q}}\Bigr) = \vert \mbox{det\1}\phi_{\s q}\vert^{\s -1}
\5 \v\Bigl(\mbox{\H{q}}\Bigr).
\ee
So the volume of the stabilizer is:
\be
\v(\kd{0}) =
\df{0}\+ \5
\v\Bigl(\mbox{\h{0}}\Bigr). 
\ee
The volume of the orbit of the group is proportional to the ghost--for--ghost
determinant $\mbox{det\2}(\phi^{\dagger}_{\s 0}
\phi^{\phantom{\dagger}}_{\s 0})$ that extracts the zero modes from the 
Faddeev--Popov ghost determinant $\mbox{det\2}(d_{\s 1}^{\dagger}
d_{\s 1}^{\phantom{\dagger}}).$
The ghost--for--ghost determinant is equal to the inverse of the volume of the 
manifold~\cite{da}:
\be
\df{0}^{\s -1}=\v(\mbox{M}).
\ee
Now we will extract the complex scaling factor $\frac{i \l{1}}{\pi}$ from the
functional determinant. Following~\cite{sen} we can write:
\be
\mbox{det}^{\prime}\1\Bigl(\frac{i \l{1}}{\pi}\2 d^{\dagger}_{\s 1}
d^{\phantom{\dagger}}_{\s 1}\Bigr)\- \5 = \5
e^{-\frac{i\pi}{4}\eta(0, \2 d_{\s 1}^{\dagger} d_{\s 1}^{\phantom{\dagger}})}
\5 \biggl(\frac{\l{1}}{\pi}\biggr)^{-\frac{1}{2}
\zeta(0, \2 d_{\s 1}^{\dagger} d_{\s 1}^{\phantom{\dagger}})}\5
\dd{1}\-.
\ee
Thus the partition function of Maxwell theory is given by:
\be
Z_{\s 1}(\lambda_{\s 1}) \5 = \5 
e^{-\frac{i\pi}{4}\eta(0, \2 d_{\s 1}^{\dagger} d_{\s 1}^{\phantom{\dagger}})}
\5 \biggl(\frac{\l{1}}{\pi}\biggr)^{-\frac{1}{2}
\zeta(0, \2 d_{\s 1}^{\dagger} d_{\s 1}^{\phantom{\dagger}})}\5
\frac{\mbox{\v(M)}\+}{\v\Bigl(\h{0}\Bigr)} \5
\frac{\dd{0}\+}{\dd{1}\+}.
\ee
We now use the fact that on odd-dimensional and two-dimensional manifolds 
there are no poles in the $\zeta$--function near $s=0$. This can be seen 
using Seeley's formula~\cite{seeley} for the $\zeta$-function of some 
Laplace-type operator $L$ on a $d$-dimensional manifold without a boundary:
\be
\zeta(s, L)=\frac{1}{\Gamma(s)}\sum\limits_{\s n=0}^{\s \infty}
\frac{A_{\s n}}{s + n - \frac{d}{2}} + \frac{J(s)}{\Gamma(s)},
\ee
where $A_{\s n}$ are the heat-kernel co-efficients and $J(s)$ is analytic. 
Then \linebreak $\zeta(0, \2 \Delta_{\s p}) = - \mbox{dim\1}\H{p}$. Using the
formula~\cite{sen}:
\be
\zeta(s, \2 d_{\s p}^{\dagger} d_{\s p}^{\phantom{\dagger}}) \5 = \5
(-1)^{\s p} \5 \sum\limits_{\s q=0}^{\s p} (-1)^{\s q} \5 
\zeta(s, \2 \Delta_{\s q}),
\ee
we finally get, modulo the phase factor:
\be \label{1}
Z_{\s 1}(\lambda_{\s 1}) \5 = \5
\lambda_{\s 1}^{-\frac{1}{2}\mbox{\scriptsize dim\1\H{0}}} \5
\frac{\mbox{\v(M)}\+}{\v\Bigl(\h{0}\Bigr)} \5
\dd{0}\+ \5 \dd{1}\-.
\ee

\vskip1cm
\noindent
Consider now the partition function of free scalar theory:
\ba
Z_{\s 0}(\lambda_{\s 0}) \5 & = & \5 
\int\limits_{\Omega^{\s 0} (M)}\!\!\!\!\mathcal{D} \varphi \5\1
e^{\2-i \2 \l{0} \int \!\! d^{\s 3}x \2 \sqrt{g} \2\2 
\partial_{\s \mu}\varphi \2 \partial^{\s \mu}\varphi} \5 = \5
\int\limits_{\Omega^{\s 0} (M)}\!\!\!\!\mathcal{D} \varphi \5\1 
e^{\2-i \2 \l{0} \int d_{\s 0}\varphi \2 \wedge\2 \ast\2 
d_{\s 0}\varphi} \nonumber \\ 
& = & \5 \int\limits_{\Omega^{\s 0} (M)}\!\!\!\!\mathcal{D} \varphi \5\1 
e^{\2-i \2 \l{0} \2 \langle d_{\s 0}\varphi, \2 d_{\s 0}\varphi \rangle}
\hskip34pt  = \5 \int\limits_{\Omega^{\s 0}(M)}\!\!\!\!\mathcal{D} \varphi 
\5\1 e^{\2-i \2 \l{0} \2 \langle \varphi, \2 d^{\dagger}_{\s 0} 
d^{\phantom{\dagger}}_{\s 0} \varphi \rangle}. \nonumber \\
\ea
We now decompose the space of all 0-forms $\Omega^{\s 0}(M)$ in a similar way:
\be
\Omega^{\s 0}(M) = 
\mbox{ker\2} d_{\s 0} \oplus (\mbox{ker\2} d_{\s 0})^{\perp}.
\ee
With this decomposition the partition function becomes:
\ba
Z_{\s 0}(\lambda_{\s 0}) &  =  &
\v(\kd{0}) \5
\int\limits_{(\mbox{\scriptsize ker\2} d_{\s 0})^{\perp}}
\!\!\!\!\mathcal{D}\varphi\5\1
e^{\2-i \2 \l{0} \2 \langle \varphi, \2 d^{\dagger}_{\s 0} 
d^{\phantom{\dagger}}_{\s 0} \varphi \rangle} \nonumber \\
& = &  
e^{-\frac{i\pi}{4}\eta(0, \2 d_{\s 0}^{\dagger} d_{\s 0}^{\phantom{\dagger}})}
\5 \biggl(\frac{\l{0}}{\pi}\biggr)^
{\frac{1}{2}\mbox{\scriptsize dim\1\H{0}}}\5
\v(\kd{0}) \5 \dd{0}\- \nonumber \\
& = & 
e^{-\frac{i\pi}{4}\eta(0, \2 d_{\s 0}^{\dagger} d_{\s 0}^{\phantom{\dagger}})}
\5 \biggl(\frac{\l{0}}{\pi}\biggr)^{\frac{1}{2}\mbox{\scriptsize dim\1\H{0}}}\5
\frac{\v\Bigl(\h{0}\Bigr)}{\mbox{\v(M)}\+} \5 \dd{0}\-\!\!.
\ea
Modulo the phase factor we have:
\be
Z_{\s 0}(\lambda_{\s 0}) \5 = \5 
\lambda_{\s 0}^{\frac{1}{2}\mbox{\scriptsize dim\1\H{0}}}\5
\frac{\v\Bigl(\h{0}\Bigr)}{\mbox{\v(M)}\+} \5 \dd{0}\-.
\ee
The product of the partition functions of the theories is:
\be \label{z1z2}
Z_{\s 0}(\lambda_{\s 0}) Z_{\s 1}(\lambda_{\s 1}) \5 = 
\5 \biggl(\frac{\lambda_{\s 0}}{\lambda_{\s 1}}\biggr)^
{\frac{1}{2}\mbox{\scriptsize dim\1\H{0}}_{\phantom{\frac{1}{2}}}} \5 \dd{1}\-.
\ee
On the other hand we have:
\be
Z_{\s 1}(\lambda_{\s 1}) \5 = \5
\lambda_{\s 1}^
{-\frac{1}{2}\mbox{\scriptsize dim\1\H{0}}} \5 
\v(\kd{1}) \5 \dd{1}^{\s -1/4} \5 \dd{1}^{\s -1/4}.
\ee
The Hodge star operator is invertible and on three--dimensional manifolds we 
have: $\dd{1}\+ = \mbox{det$^{\prime}$\1}(\ast \1 d_{\s 1})$. 
(For these operators the multiplicative anomaly vanishes.) Thus (modulo a 
phase factor):
\be
Z_{\s 1}(\lambda_{\s 1}) \5 = \5
\lambda_{\s 1}^
{-\frac{1}{2}\mbox{\scriptsize dim\1\H{0}}} \5
\dd{1}^{\s -1/4} \5
\v\Bigl(\mbox{ker\1}(\ast d_{\s 1})\Bigr) \5
\mbox{det$^{\prime}$\1}(\ast \1 d_{\s 1})\-.
\ee
The last two factors in this formula are exactly the partition function of 
Chern--Simons theory $Z_{\s CS}$. The partition function of Chern--Simons 
theory is a topological invariant (modulo a phase factor~\cite{sen}), given by 
the Ray--Singer torsion of the manifold~\cite{s1}:
\be
Z_{\s CS}(\lambda_{\s CS}) \5 = 
\lambda_{\s CS}^{-\frac{1}{2}\mbox{\scriptsize dim\1\H{0}}}
\5 \tau_{\mbox{\tiny RS}}\+(\mbox{M}).
\ee 
Therefore:
\be \label{z1/zcs}
\biggl(\frac{Z_{\s 1}(\lambda_{\s 1})}{Z_{\s CS}(\lambda_{\s CS})}\biggr)^{2} 
\5 = \5 \lambda_{\s 1}^{-\mbox{\scriptsize dim\1\H{0}}} \5 \dd{1}\-.
\ee
\renewcommand{\thefootnote}{$\ast$}
Dividing (\ref{z1/zcs}) by (\ref{z1z2}) we get\footnote{
After this work was completed, our attention was kindly drawn by A. Schwarz
to~\cite{st} where the ratio $Z_{\s k-1}/Z_{\s m-k-1}$ (where $m$ is the 
dimension of the manifold) is expressed as the Ray--Singer torsion. The 
difference between our work and~\cite{st} is in the following. In~\cite{st}, 
the initial considerations are for the case when there are no zero modes of 
the Laplace operators $\Delta_{\s k}$ (acting on k-forms). When these zero 
modes are absent, it is rather obvious that the quotient 
$Z_{\s k-1}/Z_{\s m-k-1}$ is the Ray--Singer torsion of the manifold. The case 
of interest appears when these zero modes are no longer neglected.
In~\cite{st} a very deep analysis is given for this case: the theory of the 
measure of the path integrals involved is developed and certain general 
results are given in this direction. In our paper we have kept these zero 
modes all along and we have shown that even with them the quotient 
($Z_{\s 1}/Z_{\s 0}$ in our case) is still given by the Ray--Singer torsion. 
In addition we have studied the scaling dependence of the models and we have 
shown the relation to \r duality. We have also given treatment on the 
physically relevant case --- interaction with external currents and 
correlation functions.}:
\be \label{result}
\frac{Z_{\s 1}(\lambda_{\s 1})}{Z_{\s 0}(\lambda_{\s 0})} \5 = 
\5 Z_{\s CS}^{\scriptstyle\2 2}\Bigl(\sqrt{\lambda_{\s 0}\lambda_{\s 1}}\5
\Bigr) \5 = \5 
(\lambda_{\s 0}\lambda_{\s 1})^{-\frac{1}{2}\mbox{\scriptsize dim\1\H{0}}}
\5 \tau_{\mbox{\tiny RS}}(\mbox{M}).
\ee
\r duality means that if the coupling constants (overall scaling factors) are 
related as $\lambda_{\s 0} = \lambda_{\s 1}^{\s -1}$ then both partition 
functions will depend on the coupling constants in the same way (one has to 
be careful, because the coupling constants are not dimensionless). The ratio 
of the partition functions (modulo an omitted phase factor) is a topological 
invariant --- the Ray--Singer torsion of the manifold. Therefore the two
theories are equivalent. For manifolds for which the Ray--Singer torsion is 
one ({\bf S}$^{\scriptstyle 3}$ for instance), the partition functions are 
equal. \\
Note that both scalar field theory and Maxwell theory are non-topological 
in three dimensions. \\
Abelian Duality 
\be
\frac{Z_{\s 1}(\lambda_{\s 1})}{Z_{\s 0}(\lambda_{\s 0})} \5 = \5 
Z_{\s CS}^{\scriptstyle\2 2}\Bigl(\sqrt{\lambda_{\s 0}\lambda_{\s 1}}\5\Bigr)
\ee
is stronger than \r duality
\be
\frac{Z_{\s 1}(\lambda_{\s 1})}{Z_{\s 0}\Bigl(\frac{1}{\lambda_{\s 1}}\Bigr)} 
\5 = \5 \tau_{\mbox{\tiny RS}}(\mbox{M})
\ee
in the sense that if the coupling constants are not related as \r, there is 
still a relation --- the ratio of the partition functions is given by the 
square of the partition function of Chern--Simons theory with coupling 
constant $\lambda_{\s CS}=\sqrt{\lambda_{\s 1}\lambda_{\s 0}}$, 
that is, by the partition function of \2 U(1) $\times$ U(1) \2 Chern--Simons 
theory with purely off-diagonal coupling (BF theory). 

\vskip1cm
\noindent
We can show the Abelian Duality by considering the following resolvent 
generated by the de Rham complex: 
\be
0 \5 \longrightarrow \5 \mbox{I}\mkern-3.5mu\mbox{R}\5
\raise9pt \hbox{$\5\2\2 \scriptstyle \phi_{\s 0}^{\s -1}$} \2
\mkern-35mu -\mkern-8.5mu
\longrightarrow
\2\1 \o{0} \5
\raise8pt \hbox{$\5 \scriptstyle d_{\s 0}$}
\mkern-25mu -\mkern-8.5mu\longrightarrow 
\2\1 \o{1} \5
\raise8pt \hbox{$\5 \scriptstyle d_{\s 1}$} 
\mkern-25mu -\mkern-8.5mu\longrightarrow 
\2\1 \kd{1} \2\1 = \2\1 \mbox{ker\2}S_{\s 1} \2\1 \longrightarrow \2\1 0.
\ee
The first cohomology group of the mainfold is trivial and we have:
\be
\v(\kd{1}) \5 = \5 \v(\i d_{\s 0}).
\ee
If we denote by $\tilde{d_{\s 0}}$ the restriction of $d_{\s 0}$ over 
$(\kd{0})^{\perp}$, then the map
\be
\tilde{d_{\s 0}}: (\kd{0})^{\perp} \longrightarrow \i d_{\s 0}
\ee
implies:
\be
\v(\i d_{\s 0}) \5 = \5 \dd{0}\+ \v\Bigl((\kd{0})^{\perp}\Bigr) \5 = \5
\dd{0}\+ \frac{\v\Bigl(\o{0}\Bigr)}{\v(\kd{0})}.
\ee
We thus get:
\be
\v(\kd{1}) \2 = \2 \v\Bigl(\o{0}\Bigr) \2 \dd{0}\+ \2 \frac{1}{\v(\kd{0})}.
\ee
We have already seen that 
\be
\v(\kd{0}) \5 = \5 \df{0}\+ \5 \v\Bigl(\h{0}\Bigr).
\ee
Therefore the partition function of Maxwell theory is given (modulo a phase
factor and with suitable normalization) by the same expression as (\ref{1}).

\vskip1cm
\noindent
To show that the Abelian Duality is a property of three dimensions only, 
consider again the de Rham complex. Due to Hodge duality, 
\5 $d_{\s m-p-1}^{\dagger} = \ast \2 d_{\s p} \2 \ast$.
\5 Therefore \5 $\dd{m-p-1} = \dd{p}$ and for even--dimensional manifolds all
determinants involving the differential operator cancel each other. In 
higher odd dimensions, it is possible to find a relation between scalar 
field theory and Maxwell theory, but there will be more determinants coming 
in from the de Rham complex, thus non-physical theories should also be 
involved.

\vskip1cm
\noindent
\r duality can be shown in a different manner --- by a duality transformation.
We will illustrate this by considering the following partition function:
\be
Z \5  =  \5 \int\limits_{\mbox{\scriptsize ker\2} d_{\s 1}}
\!\!\!\!\mathcal{D}A\5\1
e^{\2 R \int \! A \2 \wedge \2 \ast \2 A}
\5 = \5 \int\limits_{\mbox{\scriptsize ker\2} d_{\s 1}}
\!\!\!\!\mathcal{D}A\5\1
e^{\2 R \2 \langle A, \2 A \rangle}.
\ee
Over the space of the kernel of the operator $d_{\s 1}$ we can locally write 
$A = d_{\s 0} \Phi$. 
Therefore $Z$ becomes the partition function of free scalar field theory:
\be
Z \5 = \5
\int\limits_{\Omega^{\s 0} (M)}\!\!\!\!\mathcal{D} \Phi \5\1 
e^{\2 R \int d_{\s 0}\Phi \2 \wedge\2 \ast\2 
d_{\s 0}\Phi}.
\ee
Alternatively, we can replace the integral over the kernel of the operator 
$d_{\s 1}$ by an integral over $\Omega^{\s 1}(M)$ and include a Lagrange
multiplier B $\Bigl(B \in \Omega^{\scriptstyle 1}(M)\Bigr)$ to keep track of 
the fact that $A$ is flat:
\be \label{here}
Z \5  =  \5 \int\limits_{\mbox{\scriptsize ker\2} d_{\s 1}}
\!\!\!\!\mathcal{D}A\5\1
e^{\2 R \int \! A \2 \wedge \2 \ast \2 A}
\5 = \5 \int\limits_{\mbox{\scriptsize $\Omega^{\s 1}(M)$}}
\!\!\!\!\mathcal{D}A \2\1 \mathcal{D}B \5\1
e^{\2 R \int \! A \2 \wedge \2 \ast \2 A \2\1 + \2\1 
\int B \2 \wedge \2 d_{\s 1} A}.
\ee
If we integrate over $A$ and absorb the resulting determinant 
$\mbox{det\2 (R1}\mkern-4mu\mbox{I)}$ in the normalization, we end up with the
partition function of Maxwell theory with coupling constant $1/R$:
\be
Z \5  =  \5 
\int\limits_{\mbox{\scriptsize $\Omega^{\s 1}(M)$}}
\!\!\!\!\mathcal{D}B \5\1
e^{\2 \frac{1}{R} \int d_{\s 1}B \2 \wedge \2 \ast \2 d_{\s 1}B}.
\ee
The same can be seen if we make a change in the 
variables in (\ref{here}) --- \emph{dualization} --- $A \longrightarrow 
A^{\prime} = A + \frac{1}{R} \ast d_{\s 1}B$. \\
With this dualization the partition function becomes:
\be
Z \5  =  \5 
\int\limits_{\mbox{\scriptsize $\Omega^{\s 1}(M)$}}
\!\!\!\!\mathcal{D}A \5\1
e^{\2 R \int \! A \2 \wedge \2 \ast \2 A}
\int\limits_{\mbox{\scriptsize $\Omega^{\s 1}(M)$}}
\!\!\!\!\mathcal{D}B \5\1
e^{\2 \frac{1}{R} \int d_{\s 1}B\2 \wedge \2 \ast \2 d_{\s 1}B}.
\ee
The integral over $A$ is Gaussian and can be absorbed in the normalization
factor. The remaining integral is the partition function of Maxwell theory. 

\vskip1cm
\noindent
Let us now include external currents $J_{\s \mu}$ in Maxwell theory and $j$ in 
scalar field theory:
\ba
Z_{\s 1} (J) & = &  
\int\limits_{\Omega^{\s 1} (M)}\!\!\!\!\mathcal{D}A \5 \1
e^{\1 -i \1 \int \!\! d^{\s 3}x \2 \sqrt{g} \2 
(F_{\s \mu \nu}\1 F^{\s \mu \nu} + q J_{\s \mu} A^{\s \mu})} = 
\int\limits_{\Omega^{\s 1} (M)}\!\!\!\!\mathcal{D}A \5 
e^{\1 -i \1 \langle A, \2\2 d^{\dagger}_{\s 1} 
d^{\phantom{\dagger}}_{\s 1} A \rangle \2 + \2 
q \2 \langle J, \2 A \rangle}, \nonumber \\
Z_{\s 0} (j) & = & 
\int\limits_{\Omega^{\s 0} (M)}\!\!\!\!\mathcal{D} \varphi \5\1
e^{\1 -i \2 \int \!\! d^{\s 3}x \2 \sqrt{g} \2 
(\partial_{\s \mu}\varphi \2 \partial^{\s \mu}\varphi + e j\varphi)} 
= \!\int\limits_{\Omega^{\s 0}(M)}\!\!\!\!\mathcal{D} \varphi 
\5 e^{\1 -i \2 \langle \varphi, \2 d^{\dagger}_{\s 0} 
d^{\phantom{\dagger}}_{\s 0} \varphi \rangle \2 + \2 
e \2 \langle j, \2\varphi \rangle}, 
\ea
where $q$ and $e$ are some charges. \\
Now extract perfect squares and perform the Gaussian integration to end up 
with:
\be
Z_{\s 0}(j) \5 = \5 Z_{\s 0}(J) \5 \tau_{\mbox{\tiny RS}}\+(\mbox{M}) \5 
exp\biggl(\2 -i \1 e \2 \langle j, \2 
\frac{\scriptstyle 1}{\scriptstyle d_{\s 0}^{\dagger}d_{\s 0}^
{\phantom{\dagger}}}
\2 j \rangle \2 \biggr) \5 
exp\biggl(\2 i \1 q \2 \langle J, \2 
\frac{\scriptstyle 1}{\scriptstyle d_{\s 1}^{\dagger}d_{\s 1}^
{\phantom{\dagger}}} \2 J \rangle \2 \biggr).
\ee
If the charges and the currents are related as:
\be
j \5 = \5 - i \2 \sqrt{\frac{q}{e}} \2 \ast d_{\s 2}^{\phantom{\dagger}}\2
(d_{\s 1}^{\1 \dagger})^{\s -1}\2 J,
\ee
then the Abelian duality will go through on the level of interactions with 
external currents. \\
Using the definition of correlation function (as a functional derivative of the
partition function with respect to the external current), we can easily relate
the correlation functions of scalar field theory, Maxwell theory and 
Chern--Simons theory. \\
We have recently shown~\cite{emo} that the partition functions of 
Maxwell--Chern--Simons theory and the self-dual model differ by the partition
function of Chern--Simons theory (thus the two theories being equivalent).
Therefore, the ratio of the partition functions of scalar field theory and
Maxwell theory is equal (modulo phase ambiguities) to the square of the ratio
of the partition functions of Maxwell--Chern--Simons theory and the self-dual 
model. We can relate the correlation functions of these five models as well. \\
Finally we would like to mention that Chern--Simons theory can
be dynamically generated from the parity--breaking part of a theory with 
massive fermions~\cite{schap} --- as gauge--invariant regularization of the 
massless fermionic determinant introduces parity anomaly given by the 
Chern--Simons theory. In this sense, the result of our paper (\ref{result}) 
implies that a theory with massive fermions (interacting with external 
currents) together with massless scalar fields (possibly iteracting with 
external currents) add up to Maxwell theory (with possible interaction with 
external currents). Thus we have a form of bosonization in three dimensions.

\end{document}